\documentclass[runningheads]{llncs}
\usepackage[T1]{fontenc}
\usepackage{graphicx}
\usepackage{algorithm}
\usepackage{algpseudocode}
\usepackage{amsmath}
\usepackage[table]{xcolor}
\usepackage{fancyvrb}
\usepackage[utf8]{inputenc}
\usepackage{colortbl}
\usepackage{array}
\usepackage{mathtools}
\usepackage{amssymb}
\DeclareUnicodeCharacter{2200}{$\forall$}
\DeclareUnicodeCharacter{21A6}{$\mapsto$}
\DeclareUnicodeCharacter{03B1}{$\alpha$}
\DeclareUnicodeCharacter{03B2}{$\beta$}
\DeclareUnicodeCharacter{2081}{\textsubscript{1}}
\DeclareUnicodeCharacter{2082}{\textsubscript{2}}
\DeclareUnicodeCharacter{2264}{$\leq$}
\usepackage{color}
\usepackage{xcolor}

\definecolor{rose}{HTML}{F4736C}
\definecolor{mint}{HTML}{57BA5D}
\definecolor{sky}{HTML}{6E9CFF}

\newcommand{\red}[1]{\textcolor{rose}{#1}}
\newcommand{\green}[1]{\textcolor{mint}{#1}}
\newcommand{\blue}[1]{\textcolor{sky}{#1}}
\begin{document}
\title{Implementing Dependent Type Theory Inhabitation and Unification}
%
%
\author{Chase Norman\orcidID{0000-0001-8954-3770} \and
Jeremy Avigad\orcidID{0000-0003-1275-315X}}
\authorrunning{C. Norman et al.}
%
\institute{Carnegie Mellon University, Pittsburgh PA
, USA \\
\email{\{chasen, avigad\}@cmu.com}}
\maketitle              

\begin{abstract}
Dependent type theory is the foundation of many modern proof assistants. Inhabitation and unification are undecidable problems that are useful for theorem proving and program synthesis. We introduce Canonical-min, a sound and complete solver for inhabitation and unification in dependent type theory, implemented in 185 lines of Lean code. This paper describes a novel implementation of dependent type theory and a monadic framework to transform the type checker into a performant solver. Finally, we introduce DTTBench, a benchmark for type inhabitation in dependent type theory.

\end{abstract}
\section{Introduction}

Automated reasoning systems algorithmically prove statements in a formal language. There are numerous automated theorem provers for first-order and higher-order logics in common use today \cite{vampire,zipperposition2,eprover,cvc5,z3}. Many modern interactive theorem provers, however, use \emph{dependent type theory} (DTT) as their formal language \cite{lean,rocq,agda}. 

There are various reasons for this choice. Interactive theorem provers allow users to define functions in a typed programming language. With DTT, this same language can be used to write proofs. A proof that $\alpha$ implies $\beta$ is expressed as a function from certificates of $\alpha$ to certificates of $\beta$, and a proof that $P(z)$ holds for all integers $z$ is expressed as a function from an integer $z$ to a certificate of $P(z)$. DTT can even express indexed families of structures like $\mathbb{R}^n$ as a function from a number $n$ to the type $\mathbb{R}^n$. 

DTT is capable of (although not limited to) constructive reasoning, allowing it to express program synthesis problems. It can even be used to define equality and other foundational notions from scratch and prove their properties. This expressivity, however, comes at a cost: it is considerably harder to automate. \emph{Type inhabitation}, the problem of finding an expression of a given type, is undecidable, and corresponds to theorem proving and program synthesis. \emph{Unification}, the problem of solving equations of terms, is also undecidable \cite{undecidable,huet}. All existing inhabitation solvers for DTT are incomplete, as they use first-order pattern unification \cite{TWELF,mimer,Agsy,AgdaAuto,sauto}, with the exception of Canonical \cite{canonical}, which conceptually implements a sound and complete algorithm by Dowek \cite{cube}.

To demystify search in DTT, we introduce Canonical-min,\footnote{Available at \url{https://github.com/chasenorman/Canonical-min}} a simplified version of Canonical that implements sound and complete inhabitation and unification for DTT. This paper contains all 185 lines of source code, excluding compiler directives and typeclass instances. In Section \ref{Checker}, we formally specify inhabitation and unification by implementing a \emph{type checker}, a program to determine whether a term is a member of a given type. Section \ref{Data} explains the data structures that represent terms and types in Canonical-min. 
Each aspect of our implementation differs heavily from existing type checkers, but is necessary to support our novel interpretation of metavariables. 
In Section \ref{Monads}, we introduce a novel monadic framework to convert the type checker into a solver for inhabitation and unification, without any code change. Section \ref{Search} connects everything together with depth first search. Finally, we introduce DTTBench,\footnote{Available at \url{https://github.com/chasenorman/DTTBench}} a benchmark of DTT inhabitation problems from the Lean standard library \cite{lean} and Mathlib \cite{mathlib}, and evaluate Canonical-min against other DTT inhabitation systems.

\section{Dependent Type Theory}
\label{Checker}

\definecolor{color02}{rgb}{0.00,0.00,1.00}\definecolor{color04}{rgb}{0.18,0.18,0.18}\definecolor{color05}{rgb}{0.40,0.30,0.11}\definecolor{color06}{rgb}{0.00,0.00,0.43}\definecolor{color07}{rgb}{0.46,0.48,0.71}\definecolor{color08}{rgb}{0.46,0.48,0.71}\definecolor{color10}{rgb}{0.07,0.46,0.27}\definecolor{color11}{rgb}{0.06,0.44,0.00}
\subsection{Terms}
A type theory defines a language, where the set of syntactically valid expressions are called \emph{terms}. Typically, this set consists of expressions from the lambda calculus, such as $(\lambda ~x ~y \mapsto y)$. We refer to $x$ and $y$ as the \emph{bindings}. 

Suppose you wish to determine whether two lambda terms are equivalent, known as judgmental or definitional equality. Two terms with distinct bindings may still considered equal if they have the same behavior, such as $(\lambda ~x \mapsto f ~x)$ and $(\lambda ~y \mapsto f ~y)$. To account for this $\alpha$-equivalence, we generate fresh variables and apply both terms to them. 

When we call the function \texttt{Term.apply}, defined in Section 3.1, to perform these applications, the resulting expressions will be $\beta$-reduced until they are in \emph{weak head normal form} (WHNF), to account for $\beta$-equivalence. A term in WHNF is a function symbol (called the \emph{head symbol}) applied to a sequence of terms (called \emph{arguments}).\footnote{Lambda expressions are also considered WHNF, but our algorithms do not deal with such terms directly because they are always applied to fresh variables.} We say the original lambda terms are equivalent if and only if these WHNF terms are equivalent. 

We implement this procedure as a function \texttt{Term.eq} in Lean. In the following code, the left arrow (\texttt{←}) indicates that a function call is monadic, which can be ignored for now, and grayed-out segments will be explained in a later section:
\begin{Verbatim}[commandchars=\\\{\}]
{\color{color02}partial}{\color{color04} }{\color{color02}def}{\color{color04} }{\color{color05}Term.eq}{\color{color04} (}{\color{color06}t1}{\color{color04} }{\color{color06}t2}{\color{color04} : Term) : }{\color{color07}Judgment}{\color{color04} := }{\color{color08}judgment}{\color{color04} }{\color{color02}do}
{\color{color02}  let}{\color{color04} }{\color{color06}vars}{\color{color04} ← fresh}
{\color{color04}  (← }{\color{color06}t1}{\color{color04}.}{\color{color06}apply}{\color{color04} }{\color{color06}vars}{\color{color04} }{\color{color07}(other t2 vars)}{\color{color04}).}{\color{color06}eq}{\color{color04} (← }{\color{color06}t2}{\color{color04}.}{\color{color06}apply}{\color{color04} }{\color{color06}vars}{\color{color04} }{\color{color07}rigid}{\color{color04})}
\end{Verbatim}
The function is labeled ``partial'' in Lean since we do not provide proof of termination. We assume that \texttt{t1} and \texttt{t2} have the same arity, and we will see later how our representation ensures that \texttt{vars} consists of the correct number of variables.

To determine whether two WHNF expressions are equal, we compare their head symbol. If these are not equal, we will throw an exception. Otherwise, we iterate through the arguments of this head symbol in both expressions, and compare them with \texttt{Term.eq}. If the head symbol and arguments are all determined to be equal, no exception is thrown.
\begin{Verbatim}[commandchars=\\\{\}]
{\color{color02}partial}{\color{color04} }{\color{color02}def}{\color{color04} }{\color{color05}WHNF.eq}{\color{color04} (}{\color{color06}w1}{\color{color04} }{\color{color06}w2}{\color{color04} : WHNF) : }{\color{color08}Judgment}{\color{color04} := }{\color{color02}do}
{\color{color02}  if}{\color{color04} }{\color{color06}w1}{\color{color04}.}{\color{color06}head}{\color{color04} == }{\color{color06}w2}{\color{color04}.}{\color{color06}head}{\color{color04} }{\color{color02}then}
{\color{color02}    for}{\color{color04} }{\color{color06}i}{\color{color04} }{\color{color02}in}{\color{color04} [}{\color{color10}0}{\color{color04}:}{\color{color06}w1}{\color{color04}.}{\color{color06}args}{\color{color04}.}{\color{color06}size}{\color{color04}] }{\color{color02}do}
{\color{color06}      w1}{\color{color04}.}{\color{color06}args}{\color{color04}[}{\color{color06}i}{\color{color04}]!.}{\color{color06}eq}{\color{color04} }{\color{color06}w2}{\color{color04}.}{\color{color06}args}{\color{color04}[}{\color{color06}i}{\color{color04}]!}
{\color{color02}  else}{\color{color04} failure }{\color{color11}-- throw an exception}
\end{Verbatim}

It will be essential that we use exceptions, as this prevents unnecessary calls to \texttt{apply} after two terms have already been determined not to be equal. We assume that \texttt{w1} and \texttt{w2} have exactly as many arguments as the arity of their head symbols, and therefore have the same number of arguments when they have the same head symbol. The exclamation point (\texttt{!}) is necessary in Lean since we do not provide proof that the array index is in bounds.

\subsection{Types}
Type theories also define a set of \emph{types}, and a set of rules for determining whether a term is a member of a given type. In DTT, the types of lambda expressions are given by \emph{dependent function types}, for example:
$$(\red{T} : \green{\texttt{Type}}) \to (\red{t} : \green{T}) \to \blue{T}$$
We call the first $\red{T}$ and $\red{t}$ the \emph{bindings}, $\green{\texttt{Type}}$ and the second $\green{T}$ the \emph{inputs}, and the last $\blue{T}$ the \emph{output}. This type represents all functions with this signature of input types and output type.

To determine whether a term \texttt{t} is a member of a type \texttt{ty}, we again account for $\alpha$-equivalence by creating fresh variables. The types of these variables will be the inputs of \texttt{ty}. When we apply \texttt{t} to these variables and reduce to WHNF, we obtain the head symbol, its type \texttt{headType}, and the arguments.

Our first obligation is to ensure that the arguments have the types the head symbol expects. To do this, we extract the inputs of \texttt{headType}. Necessarily, this operation will require the arguments, as \texttt{headType} is a dependent function type. Then, we recursively type check each argument against the corresponding input.

Our final obligation is to ensure that the output of \texttt{headType} matches the output of \texttt{ty}. As we are working with dependent types, the output of \texttt{headType} may depend on the arguments, and the output of \texttt{ty} may depend on the fresh variables. We will observe in the following section that the output of a type is a term, so we use \texttt{Term.apply} to substitute the bindings with these expressions. We then check that these are equal with \texttt{WHNF.eq}, completing the type checker. Again, grayed-out parts can be ignored for now. 
\begin{Verbatim}[commandchars=\\\{\}]
{\color{color02}partial}{\color{color04} }{\color{color02}def}{\color{color04} }{\color{color05}check}{\color{color04} (}{\color{color06}t}{\color{color04} : Term) (}{\color{color06}ty}{\color{color04} : Typ) : }{\color{color08}Judgment}{\color{color04} := }{\color{color08}judgment}{\color{color04} }{\color{color02}do}
{\color{color02}  let}{\color{color04} }{\color{color06}vars}{\color{color04} ← fresh }{\color{color06}ty}
{\color{color02}  let}{\color{color04} ⟨}{\color{color06}\_head}{\color{color04}, }{\color{color06}headType}{\color{color04}, }{\color{color06}args}{\color{color04}⟩ ← }{\color{color06}t}{\color{color04}.}{\color{color06}apply}{\color{color04} }{\color{color06}vars}
{\color{color02}  let}{\color{color04} }{\color{color06}inputs}{\color{color04} := }{\color{color06}headType}{\color{color04}.}{\color{color06}get!}{\color{color04}.}{\color{color06}inputs}{\color{color04} }{\color{color06}args}
{\color{color02}  for}{\color{color04} }{\color{color06}i}{\color{color04} }{\color{color02}in}{\color{color04} [}{\color{color10}0}{\color{color04}:}{\color{color06}args}{\color{color04}.}{\color{color06}size}{\color{color04}] }{\color{color02}do}
{\color{color04}    check }{\color{color06}args}{\color{color04}[}{\color{color06}i}{\color{color04}]! }{\color{color06}inputs}{\color{color04}[}{\color{color06}i}{\color{color04}]!}
{\color{color04}  (← }{\color{color06}headType}{\color{color04}.}{\color{color06}get!}{\color{color04}.}{\color{color06}output}{\color{color04}.}{\color{color06}apply}{\color{color04} }{\color{color06}args}{\color{color04} }{\color{color07}(other ty.output vars)}{\color{color04}).}{\color{color06}eq}
{\color{color04}    (← }{\color{color06}ty}{\color{color04}.}{\color{color06}output}{\color{color04}.}{\color{color06}apply}{\color{color04} }{\color{color06}vars}{\color{color04} }{\color{color07}rigid}{\color{color04})}
\end{Verbatim}
We assume that \texttt{t} and \texttt{ty} have the same arity.

\section{Data Structures}
\label{Data}
\subsection{Terms}
\label{Terms}

Our representation of lambda expressions decomposes them into three parts: a sequence of \emph{bindings}, a \emph{head}, and a spine of \emph{arguments}. Unlike WHNF terms, which must have a variable head symbol, we allow the head of a term to be a term itself for representing reducible expressions. Observe how this decomposition behaves on the following terms:
\[
\hbox to \textwidth{$
\hfil
\lambda\underbracket{\vphantom{()} \red{f\,x}}\mapsto
\underbracket{\vphantom{()} \green{f}}~
\underbracket{\vphantom{()} \blue{x\,x}}
\hfil
\underbracket{\vphantom{()} \red{\_}}~
\underbracket{\vphantom{()} \green{(\lambda x \mapsto x)}}~
\underbracket{\vphantom{()} \blue{(\lambda y \mapsto y)}}
\hfil
\underbracket{\vphantom{()} \red{\_}}~
\underbracket{\vphantom{()} \green{a}}~
\underbracket{\vphantom{()} \blue{\_}}
\hfil
$}
\]
This decomposition is distinct from the standard inductive definition of terms (with constructors for variables, applications, and lambda bindings) in that it has only a single constructor, improving uniformity. Furthermore, it keeps bindings and arguments in contiguous blocks with efficient random access.

We will encode terms using exclusively \emph{de Bruijn indices} \cite{debruijn}. We store an index for the head and references to the arguments in a structure called the \emph{assignment}. The index is composed of two numbers, \texttt{debruijn} and \texttt{index}, which will be explained shortly. 

We intend for terms to be mutable, so we use references to assignments, which we call \emph{metavariables}. Below is the definition of \texttt{Assignment} and the header of the function to access the assignment from the reference \texttt{MVar}:
\begin{Verbatim}[commandchars=\\\{\}]
{\color{color02}structure}{\color{color04} }{\color{color05}Assignment}{\color{color04} }{\color{color02}where}
{\color{color04}  debruijn: Nat}
{\color{color04}  index: Nat}
{\color{color04}  args: Array MVar}

{\color{color02}partial}{\color{color04} }{\color{color02}def}{\color{color04} }{\color{color05}MVar.assignment}{\color{color04} (}{\color{color06}m}{\color{color04} : MVar)}
{\color{color07}  (rigid : ConstraintM Bool)}{\color{color04} : }{\color{color07}SearchM}{\color{color04} Assignment := }{\color{color02}do}{\color{color04} ...}
\end{Verbatim}
Notably, the assignment does not store information about the bindings. We do not need to store the names of the bindings, since we are using de Bruijn indices (a similar trick is used for the locally nameless representation \cite{nameless}), and we do not need to store the number of bindings, since all terms have exactly one block of bindings. 

To associate an index with the head, we maintain a bank of all head variables and head terms in the local context, called the \emph{explicit substitution} (ES).  An ES is a linked list of blocks, where each is either a block of variables (with a block identifier, and optionally with types) or a block of terms. 

Terms and variables will always be added to the ES directly as blocks, allowing for efficient and uniform behavior regardless of block size. The \texttt{debruijn} field of \texttt{MVar} is the position of a block in the linked list, and the \texttt{index} field is the position of the head in this block. Since blocks can be added to the ES, the meaning of \texttt{debruijn} is dependent on its depth in the term. The ES is a linked list because it must be a fully persistent data structure. Our use of blocks will decrease the length of the linked list, improving access times. 

A term is defined as a metavariable, and an ES. 
\par\medskip
\noindent
\hbox to \textwidth{%
\hfil
\begin{minipage}[t]{0.3\textwidth}
\begin{Verbatim}[commandchars=\\\{\}]
{\color{color02}structure}{\color{color04} }{\color{color05}Term}{\color{color04} }{\color{color02}where}
{\color{color04}  mvar: MVar}
{\color{color04}  es: List Block}
\end{Verbatim}
\end{minipage}%
\hfil
\hfil
\begin{minipage}[t]{0.5\textwidth}
\begin{Verbatim}[commandchars=\\\{\}]
{\color{color02}inductive}{\color{color04} }{\color{color05}Block}{\color{color04} }{\color{color02}where}
{\color{color04}\textbar{} vars: Nat → Option Types → Block}
{\color{color04}\textbar{} terms: Terms → Block}
\end{Verbatim}
\end{minipage}%
\hfil
}
\par\medskip
It is essential that we separate the ES from the de Bruijn indices, since metavariables will have indices that are mutable, and may even be unassigned. To properly resolve the head after changes in index, and to perform $\beta$-reduction on terms with unassigned indices, we require the bank of head expressions. 


The type \texttt{Terms} is an efficient way to represent an array of terms that all share an ES. The term at index $i$ is defined as the metavariable at index $i$ paired with the ES:
\begin{Verbatim}[commandchars=\\\{\}]
{\color{color02}structure}{\color{color04} }{\color{color05}Terms}{\color{color04} }{\color{color02}where}
{\color{color04}  mvars: Array MVar}
{\color{color04}  es: List Block}

{\color{color02}def}{\color{color04} }{\color{color05}Terms.size}{\color{color04} (}{\color{color06}ts}{\color{color04} : Terms) : Nat := }{\color{color06}ts}{\color{color04}.}{\color{color06}mvars}{\color{color04}.}{\color{color06}size}
\end{Verbatim}
A WHNF term consists of a head variable, an optional type for that head variable, and arguments which are terms. A variable is uniquely identified by the identifier of the block it comes from, and the index into that block.
\par\medskip
\noindent
\hbox to \textwidth{%
\hfil
\begin{minipage}[t]{0.3\textwidth}
\begin{Verbatim}[commandchars=\\\{\}]
{\color{color02}structure}{\color{color04} }{\color{color05}WHNF}{\color{color04} }{\color{color02}where}
{\color{color04}  head: Var}
{\color{color04}  type: Option Typ}
{\color{color04}  args: Terms}
\end{Verbatim}
\end{minipage}%
\hfil
\hfil
\begin{minipage}[t]{0.3\textwidth}
\begin{Verbatim}[commandchars=\\\{\}]
{\color{color02}structure}{\color{color04} }{\color{color05}Var}{\color{color04} }{\color{color02}where}
{\color{color04}  blockId: Nat}
{\color{color04}  index: Nat}
\end{Verbatim}
\end{minipage}%
\hfil
}
\par\medskip
Now, we put everything together to define \texttt{apply}. To apply a term \texttt{t} to a block of arguments \texttt{args}, we begin by extending the ES of \texttt{t} to contain \texttt{args} at de Bruijn index 0, substituting for the bindings of \texttt{t}. Next, we access the assignment \texttt{assn} of \texttt{t.mvar} and index into the ES at \texttt{assn.debruijn}. If the block at this position consists of variables, we have reached WHNF. The head symbol is the variable from that block at index \texttt{assn.index}, and the arguments are \texttt{assn.args} under the extended ES. If the variables are typed, the type of the head symbol is the type at index \texttt{assn.index}.

Otherwise, if the block consists of terms, we must $\beta$-reduce. We access the term at index \texttt{assn.index} and apply it to \texttt{assn.args} under the extended ES.

\begin{Verbatim}[commandchars=\\\{\}]
{\color{color02}partial}{\color{color04} }{\color{color02}def}{\color{color04} }{\color{color05}Term.apply}{\color{color04} (}{\color{color06}t}{\color{color04} : Term) (}{\color{color06}args}{\color{color04} : Block)}
{\color{color07}  (rigid : ConstraintM Bool := pure false)}{\color{color04} : }{\color{color08}SearchM}{\color{color04} WHNF := }{\color{color02}do}
{\color{color02}  let}{\color{color04} }{\color{color06}es}{\color{color04} := }{\color{color06}args}{\color{color04} :: }{\color{color06}t}{\color{color04}.}{\color{color06}es}
{\color{color02}  let}{\color{color04} }{\color{color06}assn}{\color{color04} ← }{\color{color06}t}{\color{color04}.}{\color{color06}mvar}{\color{color04}.}{\color{color06}assignment}{\color{color04} }{\color{color08}rigid}
{\color{color02}  let}{\color{color04} }{\color{color06}args}{\color{color04} : Terms := \{ mvars := }{\color{color06}assn}{\color{color04}.}{\color{color06}args}{\color{color04}, }{\color{color06}es}{\color{color04} \}  }
{\color{color02}  match}{\color{color04} }{\color{color06}es}{\color{color04}[}{\color{color06}assn}{\color{color04}.}{\color{color06}debruijn}{\color{color04}]! }{\color{color02}with}
{\color{color04}  \textbar{} .vars }{\color{color06}blockId}{\color{color04} }{\color{color06}types}{\color{color04} =\texttt{>} pure \{      }
{\color{color04}      head := \{ }{\color{color06}blockId}{\color{color04}, index := }{\color{color06}assn}{\color{color04}.}{\color{color06}index}{\color{color04} \},      }
{\color{color04}      type := }{\color{color06}types}{\color{color04} \texttt{>}\texttt{>}= (\textperiodcentered{}[}{\color{color06}assn}{\color{color04}.}{\color{color06}index}{\color{color04}]!), }{\color{color06}args}
{\color{color04}    \}  }
{\color{color04}  \textbar{} .terms }{\color{color06}ts}{\color{color04} =\texttt{>} }{\color{color06}ts}{\color{color04}[}{\color{color06}assn}{\color{color04}.}{\color{color06}index}{\color{color04}]!.}{\color{color06}apply}{\color{color04} }{\color{color06}args}{\color{color04} }{\color{color07}rigid}
\end{Verbatim}
Note that apply does not create new metavariables or assignments.

\subsection{Types}
Just like \texttt{Term}, \texttt{Typ} consists of a de-Bruijn-indexed type-metavariable and an ES. \texttt{Types} is defined like \texttt{Terms}, with an array of type-metavariables that share an ES. The type at index $i$ is defined as the type-metavariable at index $i$ paired with the ES.
\par\medskip
\noindent
\hbox to \textwidth{%
\hfil
\begin{minipage}[t]{0.3\textwidth}
\begin{Verbatim}[commandchars=\\\{\}]
{\color{color02}structure}{\color{color04} }{\color{color05}Typ}{\color{color04} }{\color{color02}where}
{\color{color04}  mvar: TypeMVar}
{\color{color04}  es: List Block}
\end{Verbatim}
\end{minipage}%
\hfil
\hfil
\begin{minipage}[t]{0.3\textwidth}
\begin{Verbatim}[commandchars=\\\{\}]
{\color{color02}structure}{\color{color04} }{\color{color05}Types}{\color{color04} }{\color{color02}where}
{\color{color04}  mvars: Array TypeMVar}
{\color{color04}  es: List Block}
\end{Verbatim}
\end{minipage}%
\hfil
}
\par\medskip
\begin{Verbatim}[commandchars=\\\{\}]
{\color{color02}def}{\color{color04} }{\color{color05}Types.size}{\color{color04} (}{\color{color06}ts}{\color{color04} : Types) : Nat := }{\color{color06}ts}{\color{color04}.}{\color{color06}mvars}{\color{color04}.}{\color{color06}size}
\end{Verbatim}
The distinction between \texttt{TypeMVar} and \texttt{MVar} is that \texttt{TypeMVar} also contains an array of \texttt{TypeMVar} for the input types, as well as an array of preferred names for the bindings (which are exclusively used for printing, and are never compared). 
\begin{Verbatim}[commandchars=\\\{\}]
{\color{color02}structure}{\color{color04} }{\color{color05}TypeMVar}{\color{color04} }{\color{color02}where}
{\color{color04}  inputs: Array TypeMVar}
{\color{color04}  output: MVar}
{\color{color04}  preferredNames: Array String}
\end{Verbatim}
Note that the output is not itself a dependent function type, as we expect functions to be uncurried. As such it is represented as a metavariable expression.

Accessing the input types of a given type requires a block of arguments, as later inputs may depend on earlier inputs. The inputs are obtained by pairing the inputs array of \texttt{TypeMVar} with the explicit substitution, extended to instantiate the bindings at de Bruijn index 0 with the block of arguments. 
\begin{Verbatim}[commandchars=\\\{\}]
{\color{color02}def}{\color{color04} }{\color{color05}Typ.inputs}{\color{color04} (}{\color{color06}ty}{\color{color04} : Typ) (}{\color{color06}args}{\color{color04} : Block) : Types :=}
{\color{color04}  \{ mvars := }{\color{color06}ty}{\color{color04}.}{\color{color06}mvar}{\color{color04}.}{\color{color06}inputs}{\color{color04}, es := }{\color{color06}args}{\color{color04} :: }{\color{color06}ty}{\color{color04}.}{\color{color06}es}{\color{color04} \}}
\end{Verbatim}
We do not structurally enforce telescoping, i.e. that the input at index $i$ should only refer to the elements of \texttt{args} at indices $j < i$. Instead, we use the same explicit substitution for all the inputs, which includes all \texttt{args}. 

The output of a type is the \texttt{output} \texttt{MVar} paired with the explicit substitution. This is a lambda term, whose bindings must be instantiated using \texttt{Term.apply}.
\begin{Verbatim}[commandchars=\\\{\}]
{\color{color02}def}{\color{color04} }{\color{color05}Typ.output}{\color{color04} (}{\color{color06}ty}{\color{color04} : Typ) : Term :=}
{\color{color04}  \{ mvar := }{\color{color06}ty}{\color{color04}.}{\color{color06}mvar}{\color{color04}.}{\color{color06}output}{\color{color04}, es := }{\color{color06}ty}{\color{color04}.}{\color{color06}es}{\color{color04} \}}
\end{Verbatim}
As an example of \texttt{Typ.output}, here is the behavior on an example type:
$$(T : \texttt{Type}) \to (t : T) \to T \quad \overset{\text{output}}{\longmapsto} \quad \lambda ~T ~t \mapsto T$$

\section{Monads}
\label{Monads}
Thus far, we have shown the implementation of a type checker for DTT. It remains to be shown how to implement inhabitation and unification procedures. In fact, the code we have shown is already capable of this, with an artful choice of \emph{monads}. A monad is a type transformer; if \texttt{m} is a monad and $\alpha$ is a type, then \texttt{m $\alpha$} is a type, representing the type of computations that return an element of $\alpha$. These computations may use different effects, depending on \texttt{m}. For instance, \texttt{Option} is a monad where \texttt{Option $\alpha$} is the type of computations returning an element of $\alpha$ or throwing an exception. 

We would like to extend our type checker to support unassigned metavariables. If an unassigned metavariable is encountered during type checking, we should produce a \emph{constraint} on that metavariable, representing the remaining checks to be performed once the metavariable is given an assignment. This will allow us to perform a backtracking search over the assignments to metavariables, and defer type-checking constraints until they can be further evaluated. 

We define \texttt{ConstraintM $\alpha$} to be the type of computations that can read from an array of assignments and return an element of $\alpha$ or throw an exception. 
\begin{Verbatim}[commandchars=\\\{\}]
{\color{color02}abbrev}{\color{color04} }{\color{color05}ConstraintM}{\color{color04} := ReaderT (Array (Option Assignment)) Option}
\end{Verbatim}
Each metavariable will have an \texttt{id} which serves as an index into this array. We say the metavariable is unassigned if the value at this index is \texttt{none}. \texttt{ReaderT} is a \emph{monad transformer}, meaning it can extend an existing monad to add the additional capability of accessing a read-only value. In this case, we extend the \texttt{Option} monad, which allows us to throw an exception with \texttt{failure} when a constraint has been violated. 

We now define a \texttt{Constraint} to be a continuation and the unassigned metavariable that prevented further progress, called \texttt{stuck}. The continuation is a computation which either throws an exception in the case that the \texttt{Constraint} has been violated, or otherwise returns an array of \texttt{Constraint} with any remaining work to be done.

\begin{Verbatim}[commandchars=\\\{\}]
{\color{color02}structure}{\color{color04} }{\color{color05}Constraint}{\color{color04} }{\color{color02}where}
{\color{color04}  continuation: ConstraintM (Array Constraint)}\footnote{}
{\color{color04}  stuck: MVar}
{\color{color07}  rigid: ConstraintM Bool}
\end{Verbatim}

\footnotetext{Actually, we use the equivalent type \texttt{Array (Option Assignment) → Option (Array Constraint)} to convince the compiler that the definition is sound.}
Next, we define the \texttt{JudgmentM} monad, and \texttt{Judgment} type for \texttt{Term.eq}, \texttt{WHNF.eq}, and \texttt{check}:
\begin{Verbatim}[commandchars=\\\{\}]
{\color{color02}abbrev}{\color{color04} }{\color{color05}JudgmentM}{\color{color04} := StateT (Nat × Array Constraint) ConstraintM}
{\color{color02}abbrev}{\color{color04} }{\color{color05}Judgment}{\color{color04} := JudgmentM Unit}
\end{Verbatim}
\texttt{JudgmentM} extends \texttt{ConstraintM} with two state variables. The first is a \texttt{Nat}, which is used to create identifiers for fresh blocks of variables. When creating fresh variables, we return a \texttt{Block.vars} with identifier equal to the state variable \texttt{n}, and the state variable is incremented. When creating typed variables, we set the \texttt{Types} of the variables to be the \texttt{inputs} of a given type \texttt{ty}.\footnote{Accessing the inputs of \texttt{ty} requires the exact variables we are trying to create. To get around this, we create a block of variables with the same ID \texttt{n}, but without types.}
\begin{Verbatim}[commandchars=\\\{\}]
{\color{color02}def}{\color{color04} }{\color{color05}fresh}{\color{color04} (}{\color{color06}ty}{\color{color04} : Option Typ := none) : JudgmentM Block :=}
{\color{color04}  modifyGet }{\color{color02}fun}{\color{color04} (}{\color{color06}n}{\color{color04}, }{\color{color06}l}{\color{color04}) =\texttt{>}    }
{\color{color04}    (.vars }{\color{color06}n}{\color{color04} (}{\color{color06}ty}{\color{color04} \texttt{>}\texttt{>}= (\textperiodcentered{}.}{\color{color06}inputs}{\color{color04} (.vars }{\color{color06}n}{\color{color04} none))), }{\color{color06}n}{\color{color04}.}{\color{color06}succ}{\color{color04}, }{\color{color06}l}{\color{color04})}
\end{Verbatim}
The second state variable of \texttt{JudgmentM} is an array of \texttt{Constraint}, which we will accumulate as the judgment is evaluated. A \texttt{Judgment} is defined as a \texttt{JudgmentM Unit}, as the only information returned from a \texttt{Judgment} is whether an exception was thrown, the current value of the \texttt{Nat}, and the accumulated \texttt{Array Constraint}.

We will now define the workhorse of the algorithm, the \emph{continuation monad}. Consider a function with return type $\beta$ which calls a function with return type $\alpha$ in the continuation monad. The caller will provide a callback function from $\alpha$ to $\beta$ to the callee, representing all the remaining code the caller intends to perform with the return value from the callee. The caller then immediately cedes control, leaving it up to the callee to decide how many times (if any) to invoke the callback. 

Observe how the following definition of \texttt{ContT m r a} accepts a callback describing what should be done with the return type \texttt{a}. Rather than returning type \texttt{a}, the computation returns the same result type \texttt{r} as the callback. 
\begin{Verbatim}[commandchars=\\\{\}]
{\color{color02}def}{\color{color04} }{\color{color05}ContT}{\color{color04} (}{\color{color06}m}{\color{color04} : }{\color{color02}Type}{\color{color04} → }{\color{color02}Type}{\color{color04}) (}{\color{color06}r}{\color{color04} }{\color{color06}a}{\color{color04} : }{\color{color02}Type}{\color{color04}) := (}{\color{color06}a}{\color{color04} → }{\color{color06}m}{\color{color04} }{\color{color06}r}{\color{color04}) → }{\color{color06}m}{\color{color04} }{\color{color06}r}
\end{Verbatim}
We define a specialized version of the continuation monad that extends \texttt{JudgmentM} with fixed result type \texttt{Unit}:
\begin{Verbatim}[commandchars=\\\{\}]
{\color{color02}abbrev}{\color{color04} }{\color{color05}SearchM}{\color{color04} := ContT JudgmentM Unit}
\end{Verbatim}

\noindent Our intention is to use this continuation monad for \texttt{MVar.assignment}. 

Recall that the \texttt{MVar.assignment} function is supposed to access the assignment of a metavariable. If the metavariable is assigned, we will directly invoke the callback with the assignment, continuing normal execution. Otherwise, we sequester the callback as a new \texttt{Constraint} in the state array, to be invoked later. To access the assignments array, \texttt{ReaderT} provides the \texttt{read} function. In the following code, \texttt{k} is the callback and \texttt{l} is the array of constraints.

\begin{Verbatim}[commandchars=\\\{\}]
{\color{color02}partial}{\color{color04} }{\color{color02}def}{\color{color04} }{\color{color05}MVar.assignment}{\color{color04} (}{\color{color06}m}{\color{color04} : MVar)}
{\color{color07}  (rigid : ConstraintM Bool)}{\color{color04} : SearchM Assignment := }{\color{color02}do}
{\color{color04}  (← read)[}{\color{color06}m}{\color{color04}.}{\color{color06}id}{\color{color04}]!.}{\color{color06}getDM}{\color{color04} }{\color{color02}fun}{\color{color04} }{\color{color06}k}{\color{color04} (}{\color{color06}n}{\color{color04}, }{\color{color06}l}{\color{color04}) =\texttt{>} }{\color{color02}return}{\color{color04} ((), }{\color{color06}n}{\color{color04}, }{\color{color06}l}{\color{color04}.}{\color{color06}push}{\color{color04} \{}
{\color{color04}    continuation := constraint (}{\color{color02}do}{\color{color04} }{\color{color06}k}{\color{color04} (← }{\color{color06}m}{\color{color04}.}{\color{color06}assignment}{\color{color04} }{\color{color07}rigid}{\color{color04})) }{\color{color06}n}{\color{color04},}
{\color{color04}    stuck := }{\color{color06}m}{\color{color04},}
{\color{color06}    rigid}
{\color{color04}  \})}
\end{Verbatim}
With this implementation, if \texttt{Term.apply} is called on a term with an unassigned \texttt{mvar}, the computation is paused and is stored in the form of a constraint on that metavariable.

The \texttt{constraint} function converts a \texttt{SearchM T} into a \texttt{ConstraintM (Array Constraint)} by passing an empty continuation, a designated \texttt{start} value for the state \texttt{Nat}, and returning only the accumulated array of \texttt{Constraint}. 
\begin{Verbatim}[commandchars=\\\{\}]
{\color{color02}def}{\color{color04} }{\color{color05}constraint}{\color{color04} (}{\color{color06}t}{\color{color04} : SearchM }{\color{color06}T}{\color{color04})}
{\color{color04}  (}{\color{color06}start}{\color{color04} : Nat) : ConstraintM (Array Constraint) := }{\color{color02}do}
{\color{color02}  return}{\color{color04} (← }{\color{color06}t}{\color{color04} (}{\color{color02}fun}{\color{color04} }{\color{color02}\_}{\color{color04} }{\color{color06}s}{\color{color04} =\texttt{>} pure ((), }{\color{color06}s}{\color{color04})) (}{\color{color06}start}{\color{color04}, \#[])).}{\color{color06}2}{\color{color04}.}{\color{color06}2}
\end{Verbatim}
Lastly, we define a conversion from \texttt{SearchM Unit} to \texttt{Judgment}, using a trivial continuation:
\begin{Verbatim}[commandchars=\\\{\}]
{\color{color02}def}{\color{color04} }{\color{color05}judgment}{\color{color04} (}{\color{color06}j}{\color{color04} : SearchM Unit) : Judgment := }{\color{color06}j}{\color{color04} pure}
\end{Verbatim}
The \texttt{judgment} function is essential to ensure that each iteration of the \texttt{for} loops in \texttt{WHNF.eq} and \texttt{check} execute independently of one another (as opposed to each iteration being the continuation of the previous), such that we can accumulate constraints from subsequent iterations even when the previous iteration is stuck. 

\section{Search}
\label{Search}

Now that our type checker can generate constraints on unassigned metavariables, we implement term enumeration via iterative deepening depth-first search. Instead of using tree height as our metric for depth, we use a custom fuel parameter, called entropy \cite{canonical}, which increases each iteration. We also limit the size of the generated terms, which we extend each iteration and run depth-first search.
\begin{Verbatim}[commandchars=\\\{\}]
{\color{color02}def}{\color{color04} }{\color{color05}iddfs}{\color{color04} [Monad }{\color{color06}m}{\color{color04}] }{\color{color07}[MonadLiftT CanonicalM m]}
{\color{color07}  (root : MVar) (cb step : m Unit)}{\color{color04} : }{\color{color06}m}{\color{color04} Unit := }{\color{color02}do}
{\color{color07}  let start ← capacity}
{\color{color02}  let}{\color{color04} }{\color{color02}mut}{\color{color04} }{\color{color06}entropy}{\color{color04} := }{\color{color10}1000}
{\color{color02}  while}{\color{color04} }{\color{color02}true}{\color{color04} }{\color{color02}do}
{\color{color04}    extend }{\color{color10}4}
{\color{color04}    dfs }{\color{color07}root}{\color{color04} }{\color{color07}cb}{\color{color04} }{\color{color07}step}{\color{color04} }{\color{color06}entropy}{\color{color04} }{\color{color07}start}
{\color{color06}    entropy}{\color{color04} := }{\color{color06}entropy}{\color{color04} * }{\color{color10}3}
\end{Verbatim}
The \texttt{dfs} function takes a \texttt{root} metavariable representing the expression that we would like to assign all unassigned metavariables within. We invoke a heuristic function \texttt{next} to select an unassigned metavariable \texttt{mvar} from the expression tree of \texttt{root}. If no such metavariable exists, we have found a solution, and we invoke a user-provided callback \texttt{cb}. 

The \texttt{next} function also predicts the entropy that will be required to solve the remaining metavariables. If this prediction exceeds the remaining entropy, we backtrack. Otherwise, we compute the array of valid head assignments for \texttt{mvar}, called the \emph{domain}. For each of these assignments, we perform the assignment, file away the generated constraints, recursively call \texttt{dfs}, and remove the constraints. Finally, we unassign \texttt{mvar}. For the purposes of allocating new metavariable identifiers, we keep track of the first unused identifier, \texttt{pos}.
\begin{Verbatim}[commandchars=\\\{\}]
{\color{color02}variable}{\color{color04} [Monad }{\color{color06}m}{\color{color04}] (}{\color{color06}root}{\color{color04} : MVar) (}{\color{color06}cb}{\color{color04} }{\color{color06}step}{\color{color04} : }{\color{color06}m}{\color{color04} Unit) }{\color{color02}in}
{\color{color02}partial}{\color{color04} }{\color{color02}def}{\color{color04} }{\color{color05}dfs}{\color{color07} [MonadLiftT CanonicalM m]}
{\color{color04}  (}{\color{color06}entropy}{\color{color04} : Float) (}{\color{color06}pos}{\color{color04} : Nat) : }{\color{color06}m}{\color{color04} Unit := }{\color{color02}do}
{\color{color06}  step}{\color{color11} -- user-provided function for logging, cancelling, etc.}
{\color{color02}  let}{\color{color04} (some (}{\color{color06}mvar}{\color{color04}, }{\color{color02}\_}{\color{color04}), }{\color{color06}predicted}{\color{color04}) ← next }{\color{color06}root}{\color{color04} \textbar{} }{\color{color06}cb}
{\color{color02}  if}{\color{color04} }{\color{color06}entropy}{\color{color04} \texttt{<} }{\color{color06}predicted}{\color{color04} }{\color{color02}then}{\color{color04} }{\color{color02}return}
{\color{color02}  let}{\color{color04} }{\color{color06}domain}{\color{color04} ← }{\color{color06}mvar}{\color{color04}.}{\color{color06}domain}{\color{color04} }{\color{color06}pos}
{\color{color02}  for}{\color{color04} (}{\color{color06}assn}{\color{color04}, }{\color{color06}constraints}{\color{color04}) }{\color{color02}in}{\color{color04} }{\color{color06}domain}{\color{color04} }{\color{color02}do}
{\color{color06}    mvar}{\color{color04}.}{\color{color06}assign}{\color{color04} }{\color{color06}assn}
{\color{color04}    process (\textperiodcentered{}.}{\color{color06}push}{\color{color04} \textperiodcentered{}) }{\color{color06}constraints}
{\color{color04}    dfs (}{\color{color06}entropy}{\color{color04} / }{\color{color06}domain}{\color{color04}.}{\color{color06}size}{\color{color04}.}{\color{color06}toFloat}{\color{color04}) (}{\color{color06}pos}{\color{color04} + }{\color{color06}assn}{\color{color04}.}{\color{color06}args}{\color{color04}.}{\color{color06}size}{\color{color04})}
{\color{color04}    process (}{\color{color02}fun}{\color{color04} }{\color{color02}\_}{\color{color04} =\texttt{>} \textperiodcentered{}.}{\color{color06}pop}{\color{color04}) }{\color{color06}constraints}
{\color{color06}  mvar}{\color{color04}.}{\color{color06}assign}{\color{color04} none}{\color{color11} -- unassign mvar}
\end{Verbatim}
The computation in \texttt{dfs} is performed in the \texttt{CanonicalM} monad. \texttt{CanonicalM} is a state monad with access to two arrays, each indexed by the \texttt{id} of a metavariable. We have \texttt{assignments} as in \texttt{ConstraintM}, and we have \texttt{constraints} which for each metavariable identifier stores an array of constraints that are stuck on that metavariable.
\begin{Verbatim}[commandchars=\\\{\}]
{\color{color02}structure}{\color{color04} }{\color{color05}CanonicalState}{\color{color04} }{\color{color02}where}
{\color{color04}  assignments: Array (Option Assignment) := \{\}  }
{\color{color04}  constraints: Array (Array Constraint) := \{\}}

{\color{color02}abbrev}{\color{color04} }{\color{color05}CanonicalM}{\color{color04} := StateM CanonicalState}
\end{Verbatim}
When assigning a metavariable, we simply modify the assignments array at the index defined by the identifier for the metavariable. This can be done with the \texttt{modify} function for \texttt{StateM}.
\begin{Verbatim}[commandchars=\\\{\}]
{\color{color02}def}{\color{color04} }{\color{color05}MVar.assign}{\color{color04} : MVar → Option Assignment → CanonicalM Unit :=}
{\color{color02}  fun}{\color{color04} }{\color{color06}m}{\color{color04} }{\color{color06}assn}{\color{color04} =\texttt{>} }{\color{color02}do}{\color{color04} modify (\{ \textperiodcentered{} }{\color{color02}with}{\color{color04} assignments[}{\color{color06}m}{\color{color04}.}{\color{color06}id}{\color{color04}] := }{\color{color06}assn}{\color{color04} \})}
\end{Verbatim}
The \texttt{process} function modifies the \texttt{constraints} array for the \texttt{stuck} metavariables of an input array of constraints \texttt{cs}. The modification function \texttt{f} is \texttt{push} when adding constraints and \texttt{pop} when removing them. 
\begin{Verbatim}[commandchars=\\\{\}]
{\color{color02}def}{\color{color04} }{\color{color05}process}{\color{color04} (}{\color{color06}f}{\color{color04} : Array Constraint → Constraint → Array Constraint)}
{\color{color04}  (}{\color{color06}cs}{\color{color04} : Array Constraint) : CanonicalM Unit := }{\color{color02}do}
{\color{color02}  for}{\color{color04} }{\color{color06}c}{\color{color04} }{\color{color02}in}{\color{color04} }{\color{color06}cs}{\color{color04} }{\color{color02}do}{\color{color04} modify }{\color{color02}fun}{\color{color04} }{\color{color06}state}{\color{color04} =\texttt{>} \{ }{\color{color06}state}{\color{color04} }{\color{color02}with}
{\color{color04}    constraints := }{\color{color06}state}{\color{color04}.}{\color{color06}constraints}{\color{color04}.}{\color{color06}modify}{\color{color04} }{\color{color06}c}{\color{color04}.}{\color{color06}stuck}{\color{color04}.}{\color{color06}id}{\color{color04} (}{\color{color06}f}{\color{color04} \textperiodcentered{} }{\color{color06}c}{\color{color04})}
{\color{color04}  \}}
\end{Verbatim}
We define \texttt{capacity} to be the size of \texttt{assignments}.
\begin{Verbatim}[commandchars=\\\{\}]
{\color{color02}def}{\color{color04} }{\color{color05}capacity}{\color{color04} : CanonicalM Nat := }{\color{color02}do}{\color{color04} pure (← get).}{\color{color06}assignments.size}
\end{Verbatim}
The \texttt{extend} function appends \texttt{none} values to assignment and empty arrays to \texttt{constraints}.
\begin{Verbatim}[commandchars=\\\{\}]
{\color{color02}def}{\color{color04} }{\color{color05}extend}{\color{color04} (}{\color{color06}size}{\color{color04} : Nat) : CanonicalM Unit := }{\color{color02}do}
{\color{color04}  modify }{\color{color02}fun}{\color{color04} }{\color{color06}state}{\color{color04} =\texttt{>} \{ }{\color{color06}state}{\color{color04} }{\color{color02}with}
{\color{color04}    assignments := }{\color{color06}state}{\color{color04}.}{\color{color06}assignments}{\color{color04} ++ replicate }{\color{color06}size}{\color{color04} none}
{\color{color04}    constraints := }{\color{color06}state}{\color{color04}.}{\color{color06}constraints}{\color{color04} ++ replicate }{\color{color06}size}{\color{color04} \#[]}
{\color{color04}  \}}
\end{Verbatim}
To understand \texttt{domain}, we need a proper definition of \texttt{MVar}. In addition to the \texttt{id}, we store the preferred names for the bindings of this lambda expression (as with \texttt{TypeMVar}), and a linked list of \texttt{TypeMVar} called \texttt{lctx}. The \texttt{lctx} will tell us the valid \texttt{debruijn} and \texttt{index} pairs for the head of this \texttt{MVar} by collecting each \texttt{TypeMVar} whose \texttt{inputs} were added to the local context.
\begin{Verbatim}[commandchars=\\\{\}]
{\color{color02}structure}{\color{color04} }{\color{color05}MVar}{\color{color04} }{\color{color02}where}
{\color{color04}  id: Nat}
{\color{color04}  lctx: List TypeMVar}
{\color{color04}  preferredNames : Array String}
\end{Verbatim}

To demonstrate the use of \texttt{MVar}, we show how to compute the domain of a metavariable \texttt{m}. The valid \texttt{debruijn} values for the assignment of \texttt{m} are the indices of a \texttt{TypeMVar} \texttt{blockType} in \texttt{m.lctx}. For that \texttt{debruijn} index, the valid \texttt{index} values are the indices of a \texttt{TypeMVar} \texttt{headType} in \texttt{blockType.inputs}. Our plan is to construct an assignment with its head defined by \texttt{debruijn} and \texttt{index}, with arguments that are fresh metavariables. First, we check if we have the capacity in the \texttt{assignments} array to create the number of metavariables expected by \texttt{headType.inputs}. If so, we create a new metavariable for each \texttt{input}, adding the bindings of \texttt{input} to the \texttt{lctx}, and choosing the \texttt{preferredNames} of the metavariable's bindings to be the \texttt{preferredNames} for \texttt{input}. 

Then, we construct this assignment and assign \texttt{m} to it. We invoke the callbacks of the constraints that were previously stuck on \texttt{m}, to obtain an array of new constraints. If no exception is thrown, the assignment is valid and we add the assignment and its constraints to the \texttt{result} array. 
\begin{Verbatim}[commandchars=\\\{\}]
{\color{color02}def}{\color{color04} }{\color{color05}run}{\color{color04} (}{\color{color06}x}{\color{color04} : ConstraintM }{\color{color06}T}{\color{color04}) : CanonicalM (Option }{\color{color06}T}{\color{color04}) := }{\color{color02}do}
{\color{color02}  return}{\color{color04} }{\color{color06}x}{\color{color04}.}{\color{color06}run}{\color{color04} (← get).}{\color{color06}assignments}

{\color{color02}def}{\color{color04} }{\color{color05}MVar.domain}{\color{color04} (}{\color{color06}m}{\color{color04} : MVar) (}{\color{color06}pos}{\color{color04} : Nat) := }{\color{color02}do}
{\color{color02}  let}{\color{color04} }{\color{color02}mut}{\color{color04} }{\color{color06}result}{\color{color04} : Array (Assignment × Array Constraint) := \#[]}
{\color{color02}  for}{\color{color04} (}{\color{color06}blockType}{\color{color04}, }{\color{color06}debruijn}{\color{color04}) }{\color{color02}in}{\color{color04} }{\color{color06}m}{\color{color04}.}{\color{color06}lctx}{\color{color04}.}{\color{color06}zipIdx}{\color{color04} }{\color{color02}do}
{\color{color02}    for}{\color{color04} (}{\color{color06}headType}{\color{color04}, }{\color{color06}index}{\color{color04}) }{\color{color02}in}{\color{color04} }{\color{color06}blockType}{\color{color04}.}{\color{color06}inputs}{\color{color04}.}{\color{color06}zipIdx}{\color{color04} }{\color{color02}do}
{\color{color02}      if}{\color{color04} }{\color{color06}pos}{\color{color04} + }{\color{color06}headType}{\color{color04}.}{\color{color06}inputs}{\color{color04}.}{\color{color06}size}{\color{color04} \texttt{>} (← capacity) }{\color{color02}then}{\color{color04} }{\color{color02}continue}
{\color{color02}      let}{\color{color04} }{\color{color06}args}{\color{color04} := }{\color{color06}headType}{\color{color04}.}{\color{color06}inputs}{\color{color04}.}{\color{color06}zipIdx}{\color{color04}.}{\color{color06}map}{\color{color04} }{\color{color02}fun}{\color{color04} (}{\color{color06}input}{\color{color04}, }{\color{color06}k}{\color{color04}) =\texttt{>} (\{        }
{\color{color04}        id := }{\color{color06}pos}{\color{color04} + }{\color{color06}k}{\color{color04}, lctx := }{\color{color06}input}{\color{color04} :: }{\color{color06}m}{\color{color04}.}{\color{color06}lctx}{\color{color04},}
{\color{color04}        preferredNames := }{\color{color06}input}{\color{color04}.}{\color{color06}preferredNames}
{\color{color04}      \} : MVar)      }
{\color{color02}      let}{\color{color04} }{\color{color06}assn}{\color{color04} : Assignment := \{ }{\color{color06}debruijn}{\color{color04}, }{\color{color06}index}{\color{color04}, }{\color{color06}args}{\color{color04} \}      }
{\color{color06}      m}{\color{color04}.}{\color{color06}assign}{\color{color04} }{\color{color06}assn}
{\color{color02}      if}{\color{color04} }{\color{color02}let}{\color{color04} some }{\color{color06}constraints}{\color{color04} ← run}
{\color{color04}        ((← get).}{\color{color06}constraints}{\color{color04}[}{\color{color06}m}{\color{color04}.}{\color{color06}id}{\color{color04}]!.}{\color{color06}flatMapM}{\color{color04} continuation)}
{\color{color02}        then}{\color{color04} }{\color{color06}result}{\color{color04} := }{\color{color06}result}{\color{color04}.}{\color{color06}push}{\color{color04} (}{\color{color06}assn}{\color{color04}, }{\color{color06}constraints}{\color{color04})}
{\color{color02}  return}{\color{color04} }{\color{color06}result}
\end{Verbatim}

\subsection{Metavariable Selection}

The \texttt{next} function, to select the next metavariable to refine and estimate the remaining work to be done, can be implemented with an arbitrary heuristic. Since we uniformly and independently process arguments in \texttt{WHNF.eq} and inputs in \texttt{check}, we are free to select unassigned metavariable in any order. In fact, the ability to assign later metavariables before their dependencies are assigned is the essential property that makes the algorithm performant in practice. The only constraint to ensure completeness is that \texttt{next} only returns \texttt{none} when all metavariables are assigned.

For this minimal implementation we use a simple heuristic. We select the rightmost metavariable in the term, unless an earlier metavariable has a constraint that is a \emph{rigid} equation, that is, an equation $\texttt{?X} \equiv M$ where $M$ is not a metavariable. We prioritize metavariables with rigid equations because they are heavily constrained, and rightmost metavariables as they generate strong unification constraints on earlier metavariables \cite{canonical}.

We define an ordering on a pair containing a metavariable and a \texttt{Bool} for whether the metavariable has a rigid constraint:
\begin{Verbatim}[commandchars=\\\{\}]
{\color{color02}instance}{\color{color04} : Ord (MVar × Bool) }{\color{color02}where}
{\color{color04}  compare }{\color{color06}a}{\color{color04} }{\color{color06}b}{\color{color04} := compare }{\color{color06}a}{\color{color04}.}{\color{color06}2}{\color{color04} }{\color{color06}b}{\color{color04}.}{\color{color06}2}
\end{Verbatim}
We choose the next metavariable to be maximum metavariable according to this ordering. Furthermore, we estimate that each metavariable multiplies the predicted entropy by 5, unless the metavariable has a rigid equation:
\begin{Verbatim}[commandchars=\\\{\}]
{\color{color02}abbrev}{\color{color04} }{\color{color05}Next}{\color{color04} := Option (MVar × Bool) × Float}
{\color{color02}partial}{\color{color04} }{\color{color02}def}{\color{color04} }{\color{color05}next}{\color{color04} (}{\color{color06}m}{\color{color04} : MVar) : CanonicalM Next := }{\color{color02}do}
{\color{color02}  if}{\color{color04} }{\color{color02}let}{\color{color04} some }{\color{color06}assn}{\color{color04} := (← get).}{\color{color06}assignments}{\color{color04}[}{\color{color06}m}{\color{color04}.}{\color{color06}id}{\color{color04}]! }{\color{color02}then}
{\color{color06}    assn}{\color{color04}.}{\color{color06}args}{\color{color04}.}{\color{color06}foldlM}{\color{color04} (init := (none, }{\color{color10}1}{\color{color04})) }{\color{color02}fun}{\color{color04} (}{\color{color06}fst}{\color{color04}, }{\color{color06}e1}{\color{color04}) }{\color{color06}s}{\color{color04} =\texttt{>} }{\color{color02}do}
{\color{color02}      let}{\color{color04} (}{\color{color06}snd}{\color{color04}, }{\color{color06}e2}{\color{color04}) ← next }{\color{color06}s}
{\color{color04}      pure (}{\color{color02}if}{\color{color04} compare }{\color{color06}fst}{\color{color04} }{\color{color06}snd}{\color{color04} == .gt }{\color{color02}then}{\color{color04} }{\color{color06}fst}{\color{color04} }{\color{color02}else}{\color{color04} }{\color{color06}snd}{\color{color04}, }{\color{color06}e1}{\color{color04} * }{\color{color06}e2}{\color{color04})}
{\color{color02}  else}
{\color{color02}    let}{\color{color04} }{\color{color06}rigid}{\color{color04} ← run ((← get).}{\color{color06}constraints}{\color{color04}[}{\color{color06}m}{\color{color04}.}{\color{color06}id}{\color{color04}]!.}{\color{color06}anyM}{\color{color04} rigid)}
{\color{color02}    return}{\color{color04} (some (}{\color{color06}m}{\color{color04}, }{\color{color06}rigid}{\color{color04}.}{\color{color06}get!}{\color{color04}), }{\color{color02}if}{\color{color04} }{\color{color06}rigid}{\color{color04}.}{\color{color06}get!}{\color{color04} }{\color{color02}then}{\color{color04} }{\color{color10}1}{\color{color04} }{\color{color02}else}{\color{color04} }{\color{color10}5}{\color{color04})}
\end{Verbatim}
To determine whether \texttt{m} has a rigid constraint, we invoke the \texttt{rigid} predicate for each of its stuck constraints. 

The \texttt{rigid} predicate was passed into the \texttt{apply} function by the type checker, and was added to the constraint by \texttt{assignment}. To facilitate this, we have the following templates for the predicate:
\begin{Verbatim}[commandchars=\\\{\}]
{\color{color02}def}{\color{color04} }{\color{color05}rigid}{\color{color04} : ConstraintM Bool := }{\color{color02}return}{\color{color04} }{\color{color02}true}
{\color{color02}def}{\color{color04} }{\color{color05}other}{\color{color04} (}{\color{color06}t}{\color{color04} : Term) (}{\color{color06}args}{\color{color04} : Block) : ConstraintM Bool :=}
{\color{color02}  return}{\color{color04} (← constraint (}{\color{color02}do}{\color{color04} }{\color{color06}t}{\color{color04}.}{\color{color06}apply}{\color{color04} }{\color{color06}args}{\color{color04} rigid) }{\color{color10}0}{\color{color04}).}{\color{color06}isEmpty}
\end{Verbatim}
Here, \texttt{rigid} indicates that the constraint is unconditionally rigid, and \texttt{other} indicates that the constraint is rigid when \texttt{t} applied with \texttt{args} does not get stuck on a metavariable. 

This concludes the implementation of inhabitation and unification. As a final performance note, we recommend only adding blocks to linked lists when the block is non-empty, which can be achieved with the following changes in \texttt{Term.apply}, \texttt{Typ.inputs}, and \texttt{domain}:
\begin{Verbatim}[commandchars=\\\{\}]
{\color{color06}es}{\color{color04} := }{\color{color02}if}{\color{color04} }{\color{color06}t}{\color{color04}.}{\color{color06}mvar}{\color{color04}.}{\color{color06}preferredNames}{\color{color04}.}{\color{color06}isEmpty}{\color{color04} }{\color{color02}then}{\color{color04} }{\color{color06}t}{\color{color04}.}{\color{color06}es}{\color{color04} }{\color{color02}else}{\color{color04} }{\color{color06}args}{\color{color04} :: }{\color{color06}t}{\color{color04}.}{\color{color06}es}
{\color{color04}...}
{\color{color04}es := }{\color{color02}if}{\color{color04} }{\color{color06}ty}{\color{color04}.}{\color{color06}mvar}{\color{color04}.}{\color{color06}inputs}{\color{color04}.}{\color{color06}isEmpty}{\color{color04} }{\color{color02}then}{\color{color04} }{\color{color06}ty}{\color{color04}.}{\color{color06}es}{\color{color04} }{\color{color02}else}{\color{color04} }{\color{color06}args}{\color{color04} :: }{\color{color06}ty}{\color{color04}.}{\color{color06}es}
{\color{color04}...}
{\color{color04}lctx := }{\color{color02}if}{\color{color04} (inputs }{\color{color06}input}{\color{color04}).}{\color{color06}isEmpty}{\color{color04} }{\color{color02}then}{\color{color04} }{\color{color06}m}{\color{color04}.}{\color{color06}lctx}{\color{color04} }{\color{color02}else}{\color{color04} }{\color{color06}input}{\color{color04} :: }{\color{color06}m}{\color{color04}.}{\color{color06}lctx}
\end{Verbatim}

\section{Results}
\label{Results}

To use Canonical-min to solve inhabitation and unification problems, create the metavariables you would like to solve, invoke the desired typing or definitional equality judgments, and then run \texttt{iddfs}. For convenience, we have implemented a Lean tactic \texttt{canonical\_min} which uses the same Lean translation layer and interface as Canonical \cite{canonical}, but uses Canonical-min to inhabit the type. To demonstrate the capabilities of this tactic, we compare against existing systems for inhabitation in DTT: Twelf, sauto, and mimer.

We introduce a new benchmark, DTTBench, consisting of 31 problems adapted from the Lean standard library and Mathlib. We selected theorems that make non-trivial use of dependent types (i.e. the proof contains a non-constant lambda expression), do not depend on definitional reduction rules beyond $\beta$-reduction, and involve common mathematical notions like equality, inequality, logical connectives, relations, and sets. We evaluated Canonical-min, Twelf, sauto, and mimer on these problems with a 60 second timeout. 
\setlength{\tabcolsep}{6pt}
\begin{figure}[ht]
\centering
\begin{tabular}{|l|c|c|c|c|}
\hline
 & Canonical-min & Twelf & sauto & mimer \\
\hline
Eq & 6/6 & 1/6 & 0/6 & 0/6 \\
\hline
Leq & 4/4 & 3/4 & 2/4 & 0/4 \\
\hline
Logic & 6/6 & 3/6 & 3/6 & 2/6 \\
\hline
Relation & 13/13 & 0/13 & 0/13 & 0/13 \\
\hline
Set & 2/2 & 1/2 & 1/2 & 0/2 \\
\hline
\hline
\textbf{Total} & \textbf{31/31} & \textbf{8/31} & \textbf{6/31} & \textbf{2/31} \\
\hline
\end{tabular}
\end{figure}

In most cases, the solvers reported failures before reaching the timeout, indicating that the intended solution was not in the search space. The differences shown here are largely qualitative: Canonical-min is complete whereas the other systems implement variants of pattern unification.

\definecolor{color02}{rgb}{0.00,0.00,1.00}\definecolor{color04}{rgb}{0.18,0.18,0.18}\definecolor{color05}{rgb}{0.40,0.30,0.11}\definecolor{color06}{rgb}{0.00,0.00,0.43}\definecolor{color07}{rgb}{0.46,0.48,0.71}\definecolor{color08}{rgb}{0.07,0.46,0.27}\definecolor{color09}{rgb}{0.06,0.44,0.00}\definecolor{color10}{rgb}{0.18,0.18,0.18}\definecolor{color11}{rgb}{0.40,0.30,0.11}\definecolor{color12}{rgb}{0.00,0.00,0.43}\definecolor{color13}{rgb}{0.06,0.44,0.00}

We now showcase Canonical-min on example Lean theorems. With the full expressivity of DTT, Canonical-min can prove fundamental properties like the transitivity of equality:
\begin{Verbatim}[commandchars=\\\{\}]
{\color{color02}theorem}{\color{color10} }{\color{color11}Eq.trans}{\color{color10} \{}{\color{color12}a}{\color{color10} }{\color{color12}b}{\color{color10} }{\color{color12}c}{\color{color10} : }{\color{color12}α}{\color{color10}\} (}{\color{color12}h₁}{\color{color10} : }{\color{color12}a}{\color{color10} = }{\color{color12}b}{\color{color10}) (}{\color{color12}h₂}{\color{color10} : }{\color{color12}b}{\color{color10} = }{\color{color12}c}{\color{color10}) : }{\color{color12}a}{\color{color10} = }{\color{color12}c}{\color{color10} :=}
{\color{color13}  -- by canonical\_min}
{\color{color10}  Eq.rec (motive := }{\color{color02}fun}{\color{color10} }{\color{color12}a\_1}{\color{color10} }{\color{color12}t}{\color{color10} =\texttt{>} }{\color{color12}a}{\color{color10} = }{\color{color12}a\_1}{\color{color10}) }{\color{color12}h₁}{\color{color10} }{\color{color12}h₂}
\end{Verbatim}
Canonical-min can reason inductively about inequalities on natural numbers:
\begin{Verbatim}[commandchars=\\\{\}]
{\color{color02}theorem}{\color{color10} }{\color{color11}succ\_le\_succ}{\color{color10} (}{\color{color12}n}{\color{color10} }{\color{color12}m}{\color{color10} : Nat) : }{\color{color12}n}{\color{color10} ≤ }{\color{color12}m}{\color{color10} → }{\color{color12}n}{\color{color10}.}{\color{color12}succ}{\color{color10} ≤ }{\color{color12}m}{\color{color10}.}{\color{color12}succ}{\color{color10} :=}
{\color{color13}  -- by canonical\_min}
{\color{color02}  fun}{\color{color10} }{\color{color12}a}{\color{color10} =\texttt{>} Nat.le.rec (motive := }{\color{color02}fun}{\color{color10} }{\color{color12}a}{\color{color10} }{\color{color12}t}{\color{color10} =\texttt{>} }{\color{color12}n}{\color{color10}.}{\color{color12}succ}{\color{color10}.}{\color{color12}le}{\color{color10} }{\color{color12}a}{\color{color10}.}{\color{color12}succ}{\color{color10})}
{\color{color10}    Nat.le.refl (}{\color{color02}fun}{\color{color10} \{}{\color{color12}m}{\color{color10}\} }{\color{color12}a}{\color{color10} }{\color{color12}a\_ih}{\color{color10} =\texttt{>} Nat.le.step }{\color{color12}a\_ih}{\color{color10}) }{\color{color12}a}
\end{Verbatim}
The following is a proof that for relations $r$ and $p$, if $f(a)$ and $f(b)$ are related by $p$ when $a$ and $b$ are related by $r$, then $f(a)$ and $f(b)$ are related by the transitive closure of $p$ when $a$ and $b$ are related by the transitive closure of $r$:
\begin{Verbatim}[commandchars=\\\{\}]
{\color{color02}theorem}{\color{color10} }{\color{color11}TransGen.lift}{\color{color10} (}{\color{color12}f}{\color{color10} : }{\color{color12}α}{\color{color10} → }{\color{color12}β}{\color{color10}) (}{\color{color12}hab}{\color{color10} : TransGen }{\color{color12}r}{\color{color10} }{\color{color12}a}{\color{color10} }{\color{color12}b}{\color{color10})}
{\color{color10}  (}{\color{color12}h}{\color{color10} : ∀ }{\color{color12}a}{\color{color10} }{\color{color12}b}{\color{color10}, }{\color{color12}r}{\color{color10} }{\color{color12}a}{\color{color10} }{\color{color12}b}{\color{color10} → }{\color{color12}p}{\color{color10} (}{\color{color12}f}{\color{color10} }{\color{color12}a}{\color{color10}) (}{\color{color12}f}{\color{color10} }{\color{color12}b}{\color{color10})) : TransGen }{\color{color12}p}{\color{color10} (}{\color{color12}f}{\color{color10} }{\color{color12}a}{\color{color10}) (}{\color{color12}f}{\color{color10} }{\color{color12}b}{\color{color10}) :=}
{\color{color13}  -- by canonical\_min}
{\color{color10}  TransGen.rec}
{\color{color10}    (motive := }{\color{color02}fun}{\color{color10} }{\color{color12}b}{\color{color10} }{\color{color12}t}{\color{color10} =\texttt{>} TransGen (}{\color{color02}fun}{\color{color10} }{\color{color12}a}{\color{color10} }{\color{color12}c}{\color{color10} =\texttt{>} }{\color{color12}p}{\color{color10} }{\color{color12}a}{\color{color10} }{\color{color12}c}{\color{color10}) (}{\color{color12}f}{\color{color10} }{\color{color12}a}{\color{color10}) (}{\color{color12}f}{\color{color10} }{\color{color12}b}{\color{color10}))}
{\color{color10}    (}{\color{color02}fun}{\color{color10} \{}{\color{color12}b}{\color{color10}\} }{\color{color12}a\_1}{\color{color10} =\texttt{>} TransGen.single (}{\color{color12}h}{\color{color10} }{\color{color12}a}{\color{color10} }{\color{color12}b}{\color{color10} }{\color{color12}a\_1}{\color{color10}))}
{\color{color10}    (}{\color{color02}fun}{\color{color10} \{}{\color{color12}b}{\color{color10} }{\color{color12}c}{\color{color10}\} }{\color{color12}a\_1}{\color{color10} }{\color{color12}a\_2}{\color{color10} }{\color{color12}a\_ih}{\color{color10} =\texttt{>} TransGen.tail }{\color{color12}a\_ih}{\color{color10} (}{\color{color12}h}{\color{color10} }{\color{color12}b}{\color{color10} }{\color{color12}c}{\color{color10} }{\color{color12}a\_2}{\color{color10})) }{\color{color12}hab}
\end{Verbatim}
Finally, we have a Cantor diagonalization argument that a function from $A$ to the powerset of $A$ cannot be surjective.
\begin{Verbatim}[commandchars=\\\{\}]
{\color{color02}theorem}{\color{color10} }{\color{color11}Cantor}{\color{color10} (}{\color{color12}f}{\color{color10} : }{\color{color12}A}{\color{color10} → (}{\color{color12}A}{\color{color10} → }{\color{color02}Prop}{\color{color10})) (}{\color{color12}f\_inv}{\color{color10} : (}{\color{color12}A}{\color{color10} → }{\color{color02}Prop}{\color{color10}) → }{\color{color12}A}{\color{color10})}
{\color{color10}  (}{\color{color12}Eq}{\color{color10} : }{\color{color02}Prop}{\color{color10} → }{\color{color02}Prop}{\color{color10} → }{\color{color02}Prop}{\color{color10}) (}{\color{color12}Not}{\color{color10} : }{\color{color02}Prop}{\color{color10} → }{\color{color02}Prop}{\color{color10}) (}{\color{color12}False}{\color{color10} : }{\color{color02}Prop}{\color{color10})}
{\color{color10}  (}{\color{color12}f\_surj}{\color{color10} : ∀ }{\color{color12}q}{\color{color10} }{\color{color12}x}{\color{color10}, }{\color{color12}Eq}{\color{color10} (}{\color{color12}f}{\color{color10} (}{\color{color12}f\_inv}{\color{color10} }{\color{color12}q}{\color{color10}) }{\color{color12}x}{\color{color10}) (}{\color{color12}q}{\color{color10} }{\color{color12}x}{\color{color10}))}
{\color{color10}  (}{\color{color12}P\_ne\_Not\_P}{\color{color10} : ∀ }{\color{color12}P}{\color{color10}, }{\color{color12}Eq}{\color{color10} }{\color{color12}P}{\color{color10} (}{\color{color12}Not}{\color{color10} }{\color{color12}P}{\color{color10}) → }{\color{color12}False}{\color{color10}) : }{\color{color12}False}{\color{color10} :=}
{\color{color13}  -- by canonical\_min}
{\color{color12}  P\_ne\_Not\_P}
{\color{color10}    (}{\color{color12}f}{\color{color10} (}{\color{color12}f\_inv}{\color{color10} }{\color{color02}fun}{\color{color10} }{\color{color12}a}{\color{color10} =\texttt{>} }{\color{color12}Not}{\color{color10} (}{\color{color12}f}{\color{color10} }{\color{color12}a}{\color{color10} }{\color{color12}a}{\color{color10})) (}{\color{color12}f\_inv}{\color{color10} }{\color{color02}fun}{\color{color10} }{\color{color12}a}{\color{color10} =\texttt{>} }{\color{color12}Not}{\color{color10} (}{\color{color12}f}{\color{color10} }{\color{color12}a}{\color{color10} }{\color{color12}a}{\color{color10})))}
{\color{color10}    (}{\color{color12}f\_surj}{\color{color10} (}{\color{color02}fun}{\color{color10} }{\color{color12}a}{\color{color10} =\texttt{>} }{\color{color12}Not}{\color{color10} (}{\color{color12}f}{\color{color10} }{\color{color12}a}{\color{color10} }{\color{color12}a}{\color{color10})) (}{\color{color12}f\_inv}{\color{color10} }{\color{color02}fun}{\color{color10} }{\color{color12}a}{\color{color10} =\texttt{>} }{\color{color12}Not}{\color{color10} (}{\color{color12}f}{\color{color10} }{\color{color12}a}{\color{color10} }{\color{color12}a}{\color{color10})))}
\end{Verbatim}
Each of these, encoded as standalone dependent function types, are included in DTTBench, and are not solved by any other system. 

\section{Future Work and Conclusion}

The type checker shown in Section \ref{Checker} and search logic shown in Section \ref{Search} are largely independent from one another. The type checker is unaware that search is being performed, and the search is unaware of how the type checker determines constraint violations. This suggests that the technique may be applied to a much broader class of languages. For instance, Canonical already has support for custom definitional reduction rules. We envision this being useful for substructural, univalent, or modal type theories. As a simpler example, using a boolean formula as the type language would yield a DPLL SAT solver \cite{dpll}, where rigid constraints are unit propagation.

We can be assured of the soundness of Canonical-min because the type checker must run to completion (including postponed constraints) without throwing an exception for the algorithm to return the result. We can be assured of completeness because we attempt every assignment for each metavariable, and use iterative deepening search. Formalizing these arguments, however, would require many proofs about the type theory, such as normalization, and about our monadic functions and data structures. We leave this to future work.

We have shown that with the proper representation of DTT and the right control flow, a sound and complete inhabitation and unification solver can be written in 185 lines of Lean code. While Canonical-min is not as capable or as optimized as Canonical \cite{canonical}, we intend for this minimal implementation to serve as a reference for understanding and analyzing the algorithm, as well as a baseline for prototyping theorem proving and program synthesis tools for DTT. 

\begin{credits}
\subsubsection{\ackname} This material is based upon work supported by the National Science Foundation Graduate Research Fellowship Program under Grant No DGE2140739, as well as NSF proposal DMS 2434614. Any opinions, findings, and conclusions or recommendations expressed in this material are those of the authors and do not necessarily reflect the views of the National Science Foundation. We thank Tesla Zhang, Harrison Grodin, and Wojciech Nawrocki for their insight.

\subsubsection{\discintname}
The authors have no competing interests to declare.
\end{credits}
%
%
%
\bibliographystyle{splncs04}
\bibliography{mybibliography}
\end{document}